\documentclass[11pt]{article}
\usepackage{amsfonts,epsfig,latexsym}

\newcommand{\R}{{\mathbb{R}}}
\newcommand{\Z}{{\mathbb{Z}}}
\newcommand{\N}{{\mathbb{N}}}

\newcommand{\beq}{\begin{equation}}
\newcommand{\eeq}{\end{equation}}
\newcommand{\bea}{\begin{eqnarray}}
\newcommand{\eea}{\end{eqnarray}}
\newcommand{\ra}{\rightarrow}

\newcommand{\cd}{\partial}

\newcommand{\osc}{C_{T,2}^+}
\newcommand{\osco}{C_{T,0}^+}
\newcommand{\om}{\Omega_{T,2}^+}
\newcommand{\omo}{\Omega_{T,0}^+}

\newcommand{\ol}{\overline}
\newcommand{\qvec}{{\bf q}}
\newcommand{\chivec}{\mbox{\boldmath{$\chi$}}}

\newtheorem{thm}{Theorem}
\newtheorem{lemma}[thm]{Lemma}

\newtheorem{defn}[thm]{Definition}

\begin{document}

\title{Breather initial profiles in chains of weakly coupled anharmonic 
oscillators}
\author{M. Haskins\thanks{E-mail: {\tt mhaskin@math.jhu.edu}} \\
Department of Mathematics, Johns Hopkins University\\
Baltimore, MD 21218, U.S.A.\\ \\
J.M. Speight\thanks{E-mail: {\tt j.m.speight@leeds.ac.uk}}\\
Department of Pure Mathematics, University of Leeds\\
Leeds LS2 9JT, England}

\date{}

\maketitle
\begin{abstract}
A systematic correlation between the initial profile of discrete breathers and their frequency is
described. The context is that of a very weakly harmonically coupled 
chain of softly anharmonic oscillators. The results are structurally stable, that is, robust under 
changes of the on-site potential and are illustrated numerically for several standard choices.
A precise genericity theorem for the results is proved.

\end{abstract}

\section{Introduction}
\label{int}

Chains of coupled anharmonic
oscillators have many applications in condensed matter and
biophysics as simple one-dimensional models of crystals or biomolecules. 
Of particular interest are the so-called ``breather''
solutions supported by such chains, that is, oscillatory solutions which
are periodic in time and exponentially localized in space \cite{review}. 
The simplest 
possible class of models, where all the oscillators are identical,
with one degree of freedom, and 
nearest neighbours are coupled by identical Hooke's law springs has been
widely studied in this context. The equation of motion for the position
$q_n(t)$ of the $n$-th oscillator ($n\in\Z$) is
\beq
\label{1}
\ddot{q}_n-\alpha(q_{n+1}-2q_{n}+q_{n-1})+V'(q_n)=0
\eeq
where $\alpha$ is the spring constant and
 $V$ is the anharmonic substrate potential, which we choose to normalize
so that $V'(0)=0$ and $V''(0)=1$.
MacKay and Aubry have proved the existence of breathers in the weak coupling
(small $\alpha$) regime of this system \cite{macaub}. They noted that in the
limit $\alpha\ra 0$, system (\ref{1}) supports a one-site breather, call it
$\qvec_0=(q_n)_{n\in\Z}$, where one site ($n=0$ say) oscillates with period
$T$ while all the others remain stationary at $0$. Using an implicit function
theorem argument they proved existence of a constant period continuation $\qvec_\alpha$ of 
this breather away from $\alpha=0$ provided $\alpha$ remains sufficiently
small and $T\notin 2\pi\Z^+$.

Having established the existence of these breathers, the question remains:
what do they look like? Of course, this problem has been intensively studied, and several 
efficient numerical schemes have been devised \cite{aubmar,greek}. The purpose of this note is
to point out that significant, systematic qualitative information about breathers in weakly coupled 
chains can be obtained from the initial {\em direction} of continuation of one-site breathers,
\beq
\qvec'_0:=\left.\frac{\cd\qvec_\alpha}{\cd\alpha}\right|_{\alpha=0}.
\eeq
Specifically, the small
$\alpha$ behaviour of breathers can be determined by approximating the
curve $\qvec_\alpha$ by its tangent line at $\alpha=0$, that is
\beq
\label{2}
\qvec_\alpha=\qvec_0+\alpha\qvec'_0+o(\alpha).
\eeq
Note that equation (\ref{2}) is a precise mathematical statement, following directly from the
MacKay-Aubry theorem, not a formal expansion.
This allows one to make a systematic study of the frequency dependence of breather initial 
profiles for several on-site potentials. The results show
a generic behaviour of alternating frequency bands of two qualitatively different
types of breather, which we call ``in-phase breathers'' and 
``anti-phase breathers''. We go on to prove that this alternating
behaviour is generic in a precise sense.

\section{The direction of continuation of one-site breathers}
\label{dir}

In the following we will assume that $V:\R\ra\R$ is twice continuously
differentiable and 
has a normalized stable equilibrium point at $0$ ($V'(0)=0$, $V''(0)=1$).
Consider the equation of motion for a particle moving in such a potential,
\beq
\label{3}
\ddot{x}+V'(x)=0
\eeq
with $\dot{x}(0)=0$. Provided $|x(0)|$ is small enough, $x(t)$ must be a
periodic oscillation. All the potentials we consider will be 
softly anharmonic
with classical frequency spectrum $(0,1)$. 

Let $x_T(t)$ denote the solution of 
(\ref{3}) with period $T>2\pi$ and $x(0)>0$ which has even time-reversal
symmetry, $x_T(-t)\equiv x_T(t)$. From this we may construct a 1-site 
breather solution of system (\ref{1}) with $\alpha=0$, call it $\qvec_0$:
\beq
\label{4}
q_{n,0}(t)=\left\{
\begin{array}{cc}
x_T(t) & n=0 \\
0 & n\neq 0.
\end{array}\right. 
\eeq
The MacKay-Aubry theorem establishes the existence of a continuation 
$\qvec_\alpha$ of this solution away from $\alpha=0$ in a suitable function
space, defined as follows.

\begin{defn} For any $n\in\N$, 
let $C^+_{T,n}$ denote the space of $n$ times 
continuously differentiable mappings $\R\ra\R$ which are $T$-periodic
and have even time reversal symmetry. Note that $C_{T,n}^+$ is a
Banach space when equipped with the uniform $C^n$ norm:
$$
|q|_n:=\sup_{t\in\R}\{|q(t)|,|\dot{q}(t)|,\ldots,|q^{(n)}(t)|\}.
$$
\end{defn}

\begin{defn} For any $n\in\N$,
$$
\Omega_{T,n}^+:=\{\qvec:\Z\ra C^+_{T,n}\,\, \mbox{\rm such that}\,\,
 ||\qvec||_n<
\infty\}
$$
where
$$
||\qvec||_n:=\sup_{m\in\Z}|q_m|_n.
$$
Note that $(\Omega_{T,n}^+,||\cdot||_n)$ is also a Banach space.
\end{defn}

The required function space is $\om$.
By construction, $\qvec_0\in\om$ and is exponentially spatially localized.

\begin{thm}(MacKay-Aubry) 
\label{th:macaub}\label{th1}
If $T\notin 2\pi\Z$ there exists $\epsilon>0$
such that for all $\alpha\in[0,\epsilon)$ there is a unique continuous
family $\qvec_\alpha\in\om$ of solutions of system (\ref{1}) at
coupling $\alpha$ with 
$\qvec_0$ as defined in (\ref{4}). These solutions are 
exponentially localized in space and the map $[0,\epsilon)\ra\om$ given by
$\alpha\mapsto\qvec_\alpha$ is $C^1$. 
\end{thm}

The idea of the proof is to define a $C^1$ mapping $F:\om\oplus\R\ra\omo$,
\beq
\label{4.25}
F(\qvec,\alpha)_m=\ddot{q}_m-\alpha(q_{m+1}-2q_m+q_{m-1})+V'(q_m),
\eeq
so that $F(\qvec,\alpha)=0$ if and only if $\qvec$ is an even $T$-periodic
 solution of system (\ref{1}). In particular, $F(\qvec_0,0)=0$ by 
construction. Using $T\notin 2\pi\Z^+$ and anharmonicity of $V$, one can show
that the partial derivative of $F$ with respect to $\qvec$ at $(\qvec_0,0)$,
$DF_{q_0}:\om\ra\omo$, is invertible ($DF_{q_0}$ is injective with 
$(DF_{q_0})^{-1}$ bounded).
Hence the implicit function theorem \cite{chobru1} applies and local 
existence and uniqueness of the $C^1$ family $\qvec_\alpha$ satisfying
\beq
\label{4.5}
F(\qvec_\alpha,\alpha)=0
\eeq
 are assured. Persistence of exponential 
localization is proved as a separate step.

The object of interest in this paper is $\qvec'_0\in\om$, the tangent vector 
to the curve $\qvec_\alpha$ at $\alpha=0$. This may be constructed by
implicit differentiation of (\ref{4.5}) with respect to $\alpha$ at $\alpha
=0$:
\beq
\label{5}
DF_{q_0}\, \qvec_0'+\left.\frac{\cd F}{\cd\alpha}\right|_{(\qvec_0,0)}=0
\qquad
\Rightarrow\qquad
\qvec_0'=-(DF_{q_0})^{-1}\, 
\left.\frac{\cd F}{\cd\alpha}\right|_{(\qvec_0,0)}.
\eeq
Both $DF_{q_0}$ and $\cd F/\cd\alpha|_{(q_0,0)}$ are easily computed from
(\ref{4},\ref{4.25}). 
Evaluating the right hand side of (\ref{5}) to find $\qvec_0'=\chivec
\in\om$ is then equivalent to solving the following infinite decoupled set of
ODEs for $\{\chi_m\in\osc:m\in\Z\}$:
\bea
\label{7}
\ddot{\chi}_0+V''(x_T)\chi_0&=&-2x_T \\
\label{8}
\ddot{\chi}_m+\chi_m&=&x_T\qquad\qquad |m|=1 \\
\label{9}
\ddot{\chi}_m+\chi_m&=&0\qquad\qquad\,\,\,\,\, |m|>1.
\eea
Recalling that $T\notin 2\pi\Z^+$ (so $\cos\notin\osc$) one sees from 
(\ref{9}) that $\chi_m\equiv 0$ for $|m|>1$. 
So one need only solve the linear boundary value
problems (\ref{7},\ref{8}) with $\dot{\chi}_m(0)=0$, $\chi_m(T)=\chi_m(0)$ 
for $m=0,1$
(clearly $\chi_{-1}\equiv\chi_1$), coupled to the nonlinear driver equation 
(\ref{3}) for
$x_T$. This is a numerically trivial task.

Having computed the initial value of the tangent vector, we may approximate
the breather initial profile for small $\alpha$ using (\ref{2}):
\beq
\label{16}
q_{m,\alpha}(0)=\left\{
\begin{array}{ccccccl}
x_T(0) & + & \alpha\chi_0(0) & + & o(\alpha) & & m=0 \\
       &   & \alpha\chi_1(0) & + & o(\alpha) & & |m|=1 \\
       &   &                 &   & o(\alpha) & & |m|>1.
\end{array}
\right.
\eeq
So to first order in $\alpha$, the continuation leaves all but the central
($m=0$) and off-central ($m=\pm 1$) sites at the equilibrium position. The
qualitative shape of the breather initial profile depends crucially on
$\chi_1(0)$ (but not $\chi_0(0)$). If $\chi_1(0)>0$, the continuation 
displaces the off-central sites from equilibrium in the same direction as
the central site. The result is a hump shaped breather in which the central
and off-central sites oscillate, roughly speaking, in phase (that is they
attain their maxima and minima simultaneously). We shall call such breathers
``in-phase breathers (IPBs).'' If $\chi_1(0)<0$, on the other hand, the
off-central sites are displaced in the opposite direction from the central 
site, resulting in a sombrero shaped initial profile. In this case, the
central and off-central sites oscillate, roughly speaking, in anti-phase
(the central site attains its maximum when the off-central sites attain their
minima, and vice-versa). We shall call such breathers ``anti-phase breathers
(APBs).'' We shall find that for a generic substrate
potential $V$, system (\ref{1}) supports both IPBs and APBs, the type
varying in bands as $T$ increases through $(2\pi,\infty)$.

\section{Numerical results}
\label{num}

The results presented in this section were generated
using a 4th order Runge-Kutta method with fixed time step $\delta t=0.01$ to solve the driver
equation (\ref{3}) for $x_T$ and the linear boundary value problems (\ref{7},\ref{8}).
The results obtained depend crucially on whether the substrate potential
has reflexion symmetry about the equilibrium position.

\begin{figure}
\centerline{\epsfysize=2truein
\epsfbox[63   200   549   589]{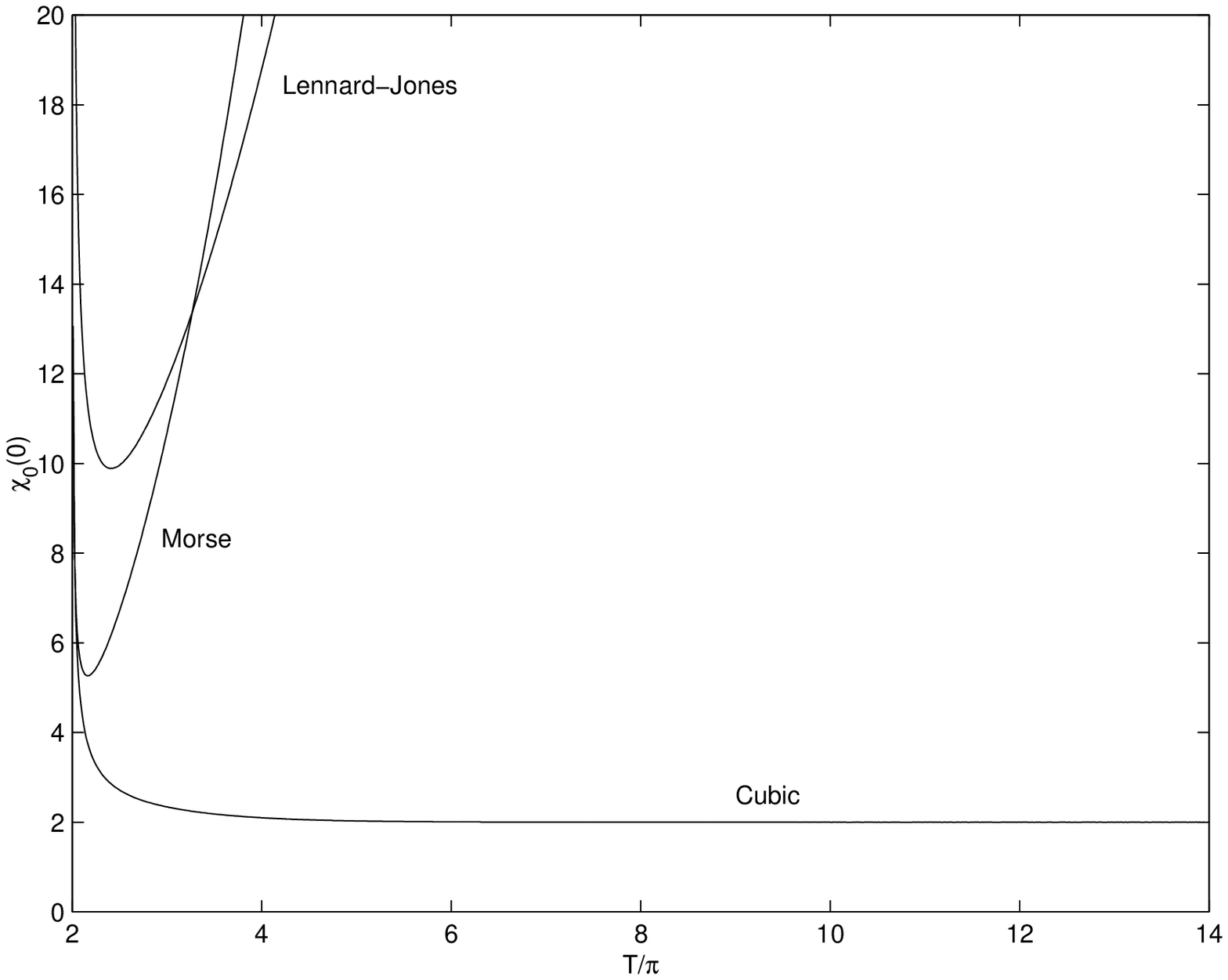}
\epsfysize=2truein
\epsfbox[63   200   549   589]{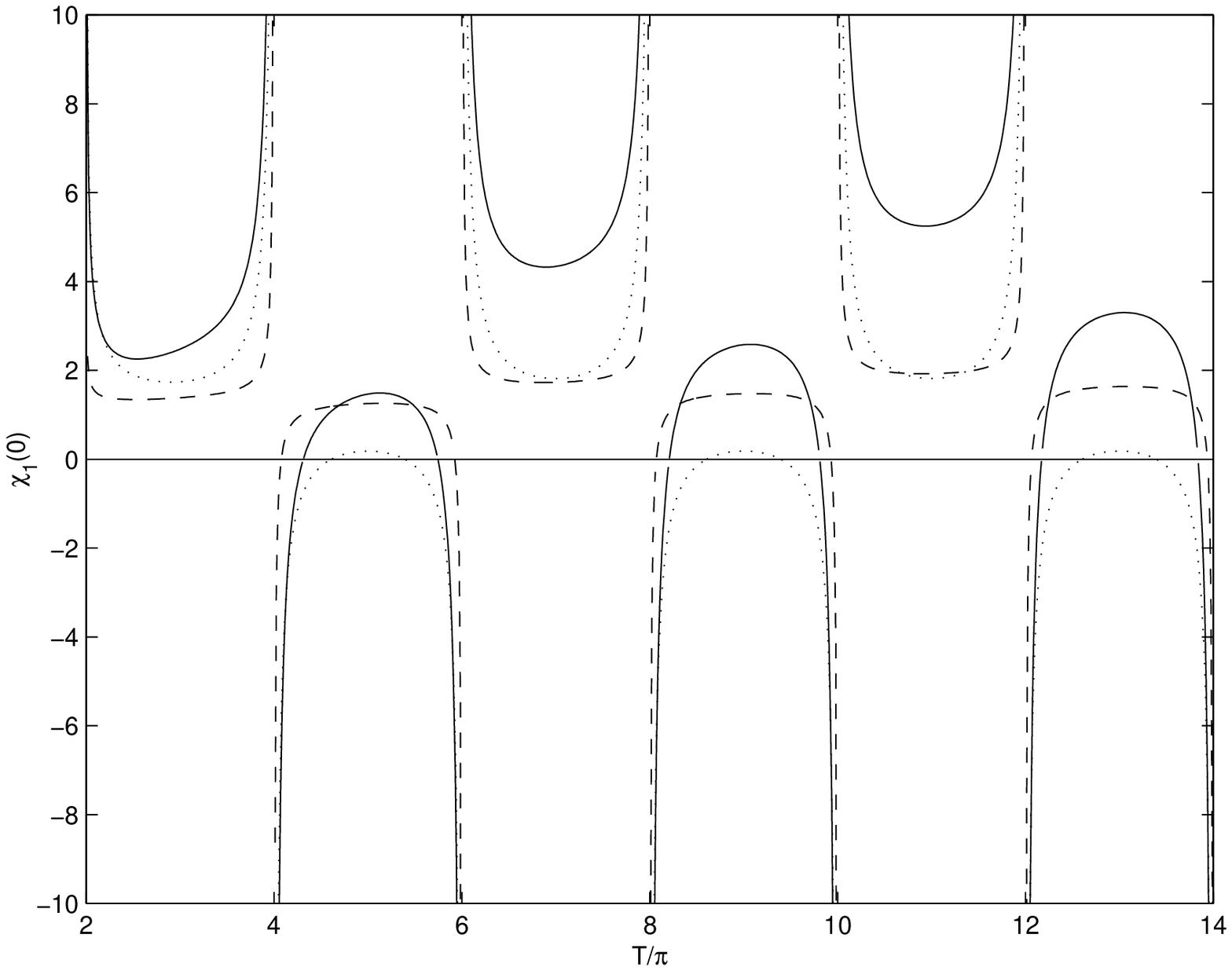}}
\caption{
Graphs of $\chi_0(0)$ (left) and $\chi_1(0)$ (right) against $T$ for various asymmetric on-site 
potentials (solid: Morse; dashed: Lennard-Jones; dotted: cubic).}
\label{fig1}
\end{figure}

The following phenomenologically standard asymmetric potentials were investigated:
\bea
\label{17}
\mbox{Morse:} &V_M(x)& =\frac{1}{2}(1-e^{-x})^2 \\
\label{18}
\mbox{Lennard-Jones:} &V_{LJ}(x)&=\frac{1}{72}\left(\frac{1}{x^{12}}-
\frac{2}{x^6}\right) \\
\label{19}
\mbox{Cubic:} &V_C(x)&=\frac{1}{2}x^2-\frac{1}{3}x^3.
\eea
Note that, although even, $V_{LJ}$ is asymmetric about its equilibrium position $x=1$.
Graphs of $\chi_0(0)$ and $\chi_1(0)$ against $T$ for all these potentials
are presented in  figure \ref{fig1}.

Since $\chi_0$ is always positive, continuation always proceeds initially by pulling the
central oscillator further away from equilibrium as $\alpha$ increases.
Note also that $\chi_0$ remains bounded as $T\ra\infty$ if the single 
oscillator solutions $x_T(t)$ remain bounded ($V_C$) but grows unbounded 
otherwise
($V_M$ and $V_{LJ}$). In all cases
it grows unbounded as $T\ra 2\pi$.

From figure \ref{fig1} we see that the sign of $\chi_1(0)$ changes as 
$T$ increases, leading to the prediction of
$T$-bands of IPBs and APBs, as explained in section \ref{dir}. The sign 
changes are clearly associated with the vertical asymptotes of the
graphs at each $T\in 2\pi\Z^+$. The presence
of such asymptotes may be understood by using Green's function techniques
to write down $\chi_1(t)$. First note that the particular integral for 
(\ref{8}) with initial
data $z(0)=\dot{z}(0)=0$ is
\beq
\label{20}
z(t)=\int_0^t\sin(t-s)x_T(s)\, ds.
\eeq
It follows from evenness of $\chi_1$ that
$
\chi_1(t)=\chi_1(0)\cos t+z(t),
$
and hence, applying the periodicity condition $\dot{\chi}_1(0)=
\dot{\chi}_1(T)$, that
\beq
\label{22}
\chi_1(0)=\frac{1}{\sin T}{\int_0^T\cos(T-t)x_T(t)\, dt}.
\eeq
Equation (\ref{22}) shows that there is a vertical asymptote at $T=2n\pi$,
with a sign change in $\chi_1(0)$, unless
\beq
\label{23}
\int_0^{2n\pi}\cos t\, x_{2n\pi}(t)\, dt=0,
\eeq
that is, unless $x_{2n\pi}(t)$ has vanishing $n$-th Fourier coefficient.
It might appear from (\ref{22}) that there should also be asymptotes at
$T=(2n+1)\pi$. This cannot be true since standard results
on continuity of solutions of ODEs with respect to initial data imply that
$\chi_1(0)$ is continuous 
for $T\notin 2\pi\Z^+$. In fact, a simple argument using
periodicity and evenness of $x_T$ demonstrates that
\beq
\label{24}
\int_0^{(2n+1)\pi}\cos t\, x_{(2n+1)\pi}(t)\, dt\equiv 0
\eeq
for all $n\in\Z^+$ and $V$. By contrast, we will prove in section \ref{gen}
that (\ref{23}) almost never holds (in a sense which will be made precise),
so that the resonant periods $T=2n\pi$ generically separate IPB bands from
APB bands. 

\begin{figure}
\centerline{\epsfysize=2truein
\epsfbox[63   200   549   589]{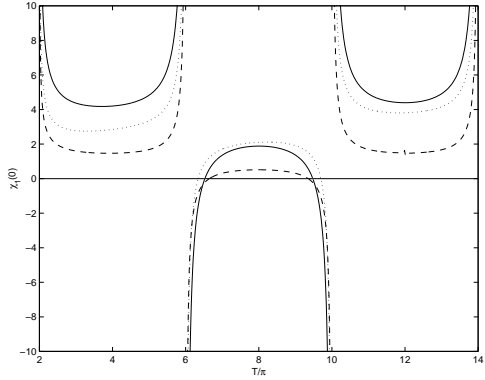}}
\noindent
\caption{
Graphs of $\chi_1(0)$ against $T$ for various symmetric on-site 
potentials (solid: Frenkel-Kontorova; dashed: quartic; dotted: Gaussian).}
\label{fig3}
\end{figure}

Turning to symmetric potentials, the following standard examples were investigated:
\bea
\mbox{Frenkel-Kontorova:} & V_{FK}(x) & =1-\cos x \\
\mbox{Quartic:}           & V_Q(x)    & =\frac{1}{2}x^2-\frac{1}{4}x^4 \\
\mbox{Gaussian:}          & V_G(x)    & =1-e^{-x^2/2}
\eea
The graphs of $\chi_0$ display identical behaviour to that found for the 
asymmetric examples, and hence are not illustrated. 
The graphs of $\chi_1(0)$ against $T$, on the other hand, 
are strikingly different from the asymmetric case (see figure \ref{fig3}). In each case, vertical
asymptotes are present at $T=2n\pi$ only if $n $ is an {\em even} positive
integer. To explain this, note that
evenness of $V$ implies that $x_T(t)$ is $T/2$ antiperiodic, that is
$x_T(t-T/2)\equiv -x_T(t)$, which in turn guarantees that equation (\ref{23})
holds whenever $n$ is odd. So alternating bands of IPBs and APBs occur for these potentials, but 
the alternation occurs half as often.

One universal feature of the graphs of $\chi_1$ (independent of symmetry) is that $\chi_1\ra\infty$
as $T\ra 2\pi$. This may easily be understood in the case where $V$ is analytic (as in all our
examples) by constructing a Linstedt expansion for $x_T(t)$ in the small amplitude regime
\cite{lin}.

\section{Genericity of numerical results}
\label{gen}

The aim of this section is to prove the generic existence of a sign change in $\chi_1$
as $T$ crosses a resonance $T\in 2\pi\Z^+$.
 In particular, we will prove that, for a given $n$, the set of 
potentials on which (\ref{23}) holds is negligibly small. 
To be precise, we will show that, for potentials in a certain
Banach space $\mathcal{P}$, the subset of potentials on which (\ref{23})
holds is locally a codimension 1 submanifold.

\begin{defn} Let $\mathcal{P}=\{V:\R\ra\R\, \mbox{such that $V$ is $C^2$ and
$||V||_2<\infty$}\}$ and $F:\osc\oplus \mathcal{P}\ra\osco$ be the mapping
$$F(q,V)=\ddot{q}+V'(q).$$
Note that $F$ is a $C^1$ mapping between Banach spaces.
\end{defn}

\begin{defn} A potential $V\in \mathcal{P}$ is 
{\rm anharmonic at $q\in\osc$} 
 if $q$ is nonconstant,  $F(q,V)=0$, and $\ker{DF_{(q,V)}}=\{0\}$, where
$DF_{(q,V)}:\osc\ra\osco$ denotes the partial derivative of $F$ at $(q,V)$.
\end{defn}

Clearly $F(q,V)=0$ means that $q$ is a solution of Newton's equation 
for motion in potential $V$. That injectivity of $DF_{(q,V)}$ is equivalent
to the standard definition of anharmonicity in Hamiltonian mechanics is 
shown, for example, in the original proof of Theorem \ref{th:macaub}
\cite{macaub}. For all the potentials we considered, $V$ is anharmonic at
$q=x_T\in\osc$ for all $T>2\pi$.

\begin{defn} A {\rm perturbation
neighbourhood of $(q,V)\in\osc\oplus \mathcal{P}$} 
is a pair $(\mathcal{U},f)$ where $\mathcal{U} \subset \mathcal{P}$ is an 
open 
set
containing $V$, and $f:\mathcal{U}\ra\osc$ is a $C^1$ map satisfying 
(a) every $W\in \mathcal{U}$ is anharmonic at $f(W)$ and (b) $f(V)=q$. Note 
that
$\mathcal{U}$ is trivially a Banach manifold.
\end{defn}

\begin{lemma} For any $V\in \mathcal{P}$ anharmonic at $q\in\osc$ there 
exists a
perturbation neighbourhood $(\mathcal{U},f)$ of $(q,V)$. The function $f$ 
is unique on 
sufficiently small $\mathcal{U}$.
\end{lemma}
{\bf Proof:} Anharmonicity implies that $DF_{(q,V)}$ is injective, and 
hence the standard solvability criterion in linear ODE theory guarantees
it is also surjective (see section 3.3.2 of \cite{hasspe} 
for details). The open mapping 
theorem then ensures
boundedness of $DF_{(q,V)}^{-1}$.
Hence $DF_{(q,V)}$ is invertible, and the implicit function theorem applied
to $F$ ensures existence of $(\mathcal{U},f)$ and local uniqueness of $f$. 
$\Box$

\begin{defn} For $T\in 2\pi\Z^+$, 
we shall say that $V\in \mathcal{P}$ is {\rm
degenerate at $q\in\osc$} if $V$ is anharmonic at $q$ and
$$\int_0^T\cos\, t\, q(t)\, dt=0.$$
\end{defn}

\begin{thm} Let $T\in 2\pi\Z^+$, 
$V_0\in \mathcal{P}$ be degenerate at $q_0\in\osc$, and 
$(\mathcal{U},f)$ be a
perturbation neighbourhood of $(q_0,V_0)$. The subset of $\mathcal{U}$ on 
which 
degeneracy persists is a codimension 1 submanifold.
\end{thm}
{\bf Proof:} Consider the $C^1$ mapping $I:\mathcal{U}\ra\R$ defined by
$$I(V)=\int_0^T\cos\, t\, [f(V)](t)\, dt$$
($I$ is $C^1$ since $f$ is $C^1$). By the Regular Value Theorem 
(see appendix)
the result follows if we establish that $0$ is a regular value of $I$, or in 
other words that $I$ is a submersion at every $V\in I^{-1}(0)$.
Since ker $DI_V$ is of finite codimension in $T_V\mathcal{U}=
\mathcal{P}$ it splits and 
hence
it suffices to show that 
$DI_V:\mathcal{P}\ra\R$ is surjective for all $V\in I^{-1}(0)$.

Let $V\in I^{-1}(0)$, $q=f(V)$.  For any $\delta V\in \mathcal{P}$, let
$\delta q_{\delta V}=-[DF_{(q,V)}^{-1}(\delta V'\circ q)]\in\osc$. In other 
words 
$\delta q_{\delta V}$ is the unique even, $T$-periodic $C^2$ solution of
$$
\ddot{\delta q}_{\delta V}+V''(q(t))\delta q_{\delta V}=-\delta V'(q(t)),
$$
whose existence follows from anharmonicity of $V$ at $q$ (invertibility
of $DF_{(q,V)}$ on $\osco$). With this notation,
$$
DI_V(\delta V)=\int_0^T\cos t\, \delta q_{\delta V}(t)\, dt
$$
and surjectivity of $DI_V$ will follow if we exhibit $\delta V \in 
\mathcal{P}$
such that $DI_V(\delta V) \ne 0$. We construct such a $\delta V$ in two 
steps.

For some $t_0\in(0,T/2)$ and $\epsilon>0$, let $b\in\osc$ be a non-negative
function with $\mbox{supp}\, b\cap[0,T/2]=[t_0-\epsilon,t_0+\epsilon]$. 
Clearly
$$
\int_0^Tb(t)\cos t\, dt =  2\int_0^{T/2}b(t)\cos t\, dt\neq 0
$$
(where $T\in 2\pi\Z^+$ has been used) 
provided we choose $t_0\notin\cos^{-1}(0)$ and $\epsilon$ sufficiently
small. Since the critical points of $q(t)$ are isolated, we may also
assume that $q$ is invertible on $[t_0-\epsilon,t_0+\epsilon]$ with
$C^2$ inverse $q^{-1}:[x_1,x_2]\ra[t_0-\epsilon,t_0+\epsilon]$. 
By its construction the function $g$ defined by
$$
g(x)=\left\{\begin{array}{cc}
-[[DF_{(q,V)}b]\circ q^{-1}](x) & x\in[x_1,x_2] \\
0 & x\notin [x_1,x_2]
\end{array}\right. 
$$ 
is in $C^0(\R)$ (since $DF$ maps to $\osco$).
This allows us to construct a function $\delta \tilde{V}(x) \in C^1(\R)$ 
such that
$\delta q_{\delta \tilde{V}}=b$ as follows. For $x \in {\rm ran}(q)$, 
$\delta \tilde{V}(x)=\int_0^xg(z)\, dz$ defines a $C^1$ function on 
ran($q$). Since 
$\delta q_{\delta \tilde{V}}$ is clearly independent of the behaviour of 
$\delta \tilde{V}$ outside
${\rm ran}(q)$, we extend $\delta \tilde{V}$ to a $C^1$ function on $\R$
with support contained in some compact set $K \supset {\rm ran}(q)$.

We cannot immediately conclude that $DI_V$ is surjective, since $\delta 
\tilde{V}$
is $C^1$ but not necessarily $C^2$. However, using a density argument it is 
straightforward
to find a $\delta V \in C^2$ close enough to $\delta \tilde{V}$ such that 
$DI_V(\delta V) \neq 0$ stills holds.
More precisely, $DI_V$ has a natural continuous extension, 
call it $\ol{DI_V}$ to $C^1$. By construction $\delta \tilde{V}$ 
satisfies $\ol{DI_V}(\delta \tilde{V})\neq 0$ and belongs to $C_K^1$ 
(the $C^1$ functions with support contained in $K$).
Since $C^2_K$ is dense in $C^1_K$, 
there exists a sequence $\delta V_m\in C^2_K\subset \mathcal{P}$ 
converging (in $C^1$ norm) to $\delta \tilde{V}$, and by continuity  
$\ol{DI_V}(\delta V_m) \neq 0$ for all $m$ sufficiently large.
$\Box$

In the above, we have chosen $\mathcal{P}=(C^2(\R),||\cdot||_2)$ as our
space of potentials. This choice was made for the sake of clarity and 
notational 
simplicity -- many other choices would work. In fact, all but two ($V_{FK}$
and $V_G$) of the example potentials considered in section \ref{num} lie,
strictly speaking, outside $\mathcal{P}$, since they are unbounded. However,
the analysis above can easily be adapted to deal with this: one simply 
replaces $\mathcal{P}$ by the affine space $\mathcal{P}_V:=\{W:||W-V||_2<
\infty\}$. 

\section{Comparison with tail behaviour}
\label{tail}

All our considerations so far have concerned the frequency dependence of
breather initial profiles in the breather {\em core}, $|m|\leq 1$. It is
interesting to compare these results with the profile behaviour in their
{\em tails}, that is, for large $|m|$. Theorem \ref{th1} guarantees that
breathers decay spatially exponentially fast, but gives no explicit 
information on the decay rate, or on sign correlations between neighbouring
sites in the tail. Such information is obtainable, at least heuristically,
by means of the Fourier analytic approach of Flach \cite{fla}. We briefly
recall this computation, adapted to our specific context.

Since the breather is spatially localized,
time periodic with period $T=2\pi/\omega$ and time reflexion
symmetric, it has a Fourier series of the form
\beq
\label{fs}
q_m(t)=\sum_{k=0}^\infty A_{m,k}\cos k\omega t,
\eeq
with $A_{m,k}\ra 0$ as $m\ra\pm\infty$ for all $k$.
Substituting (\ref{fs}) into (\ref{1}) one obtains a nonlinear algebraic
system for the coefficients $A_{m,k}$. Given the spatial decay criteria one
expects these to be close, for large $|m|$ to solutions of the linearized
algebraic system. This linearization decouples into a separate 2nd order
linear difference equation for each mode $k$, namely
\beq
\label{de}
A_{m+1,k}-\left[2+\frac{1-k^2\omega^2}{\alpha}\right]A_{m,k}+A_{m-1,k}=0,
\eeq
whose general solution lies in the span of $\mu_k^m$, $\mu_k^{-m}$ where
$\mu_k$ is a root of
\beq
\label{poly}
\mu^2-\xi_k\mu+1=0,\qquad \xi_k:=2+\frac{1-k^2\omega^2}{\alpha}
\eeq
with $|\mu_k|\leq 1$. The condition $A_{m,k}\ra 0$ implies $\mu_k$ is real
and $|\mu_k|<1$. Hence the coefficient $\xi_k$ in (\ref{poly}) must lie
outside $[-2,2]$ for all $k$, so $k\omega\notin [1,\sqrt{1+4\alpha}]$: every
harmonic of $\omega$ must lie outside the phonon frequency band. Given this,
all Fourier modes decay exponentially like $\mu_k^{-|m|}$.

In order to understand the tail behaviour, one must identify the dominant 
harmonic for large $|m|$, that is, the harmonic with slowest decay. 
Algebraically we seek that $k$, call it $k_d$ say, which has $|\mu_k|$ 
closest to 1.\, Thinking of $\mu_k$ as the preimage $f^{-1}(\xi_k)$ where
$f:(-1,1)\backslash\{0\}\ra\R$, $f(\mu)=\mu+\mu^{-1}$, one sees that this is 
the $k$ for which $|\xi_k|$ is closest to $2$.\, If $\xi_k>2$ ($k\omega<1$,
below the phonon band), then
\beq
\alpha|\xi_k-2|=1-(k\omega)^2,
\eeq
while if $\xi_k<-2$ ($k\omega>\sqrt{1+4\alpha^2}$, above the phonon band),
then
\beq
\alpha|\xi_k+2|=(1+4\alpha)-(k\omega)^2.
\eeq
Hence, the dominant harmonic is the one lying closest to the phonon band
{\em in squared frequency space.} If $k_d\omega$ lies below the band, then
$0<\mu_{k_d}<1$, and the tail has spatially uniform sign -- the analogue for
tail behaviour of an IPB.\, If $k_d\omega$ lies above the band, then
$-1<\mu_{k_d}<0$ so the breather has an alternating tail, the tail analogue
of an APB. 

\begin{figure}
\centerline{\epsfysize=2truein
\epsfbox[63   200   549   589]{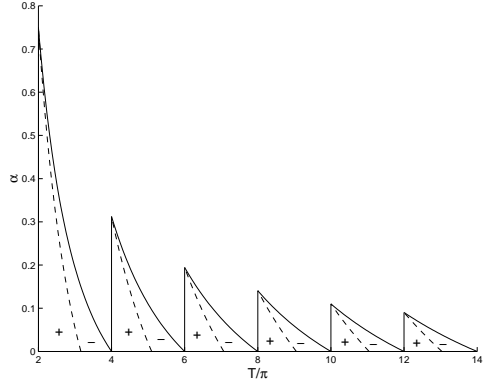}}
\noindent
\caption{
The breather existence domain for generic choice of substrate, partitioned
into IPB-tail and APB-tail sectors, denoted $+$ and $-$ respectively.}
\label{fig4}
\end{figure}

It is straightforward to partition the breather existence domain in
$(T,\alpha)$ parameter space into IPB and APB subdomains according to this
analysis, as shown in figure \ref{fig4}. In the period window $T\in[2\pi k,
2\pi(k+1)]$, one site breathers may be continued vertically from $\alpha=0$
at most until the phonon band captures the $(k+1)$th harmonic,
$(k+1)\omega=\sqrt{1+4\alpha}$, giving the upper bounding curve
\beq
T_{ex}(\alpha)=2\pi\frac{(k+1)}{\sqrt{1+4\alpha}}.
\eeq
This is the jagged upper curve depicted in figure \ref{fig4}. Within the
existence domain below this curve, the transition between IPB and
APB occurs where the pair $k\omega,(k+1)\omega$ straddling the phonon
band become equidistant (in squared frequency space) from that band:
\bea
1-k^2\omega^2&=&(k+1)^2\omega^2-(1+4\alpha)\nonumber \\
\Rightarrow
T_{trans}(\alpha)&=&2\pi\sqrt{\frac{k^2+(k+1)^2}{2+4\alpha}}.
\eea
These boundaries are depicted as dashed curves in figure \ref{fig4}. For
fixed $\alpha$, the transition with increasing $T$ across each boundary is 
always from IPB to APB, denoted $+$ and $-$ respectively in the diagram.

Comparing this picture with figure \ref{fig1} (which should be thought of as
valid in a thin strip of the $(T,\alpha)$ plane  containing the $T$ axis)
one sees that there will be many period intervals in which the core IPB/APB
classification conflicts with the tail classification. In defence of our
choice of terminology, based on the core behaviour, 
we note that breathers are interesting precisely because they are
{\em localized}
modes whose tails are exponentially small, so the core behaviour should be
regarded as more significant than the tail behaviour. It is also worth noting
two similarities between the core and tail analyses at small $\alpha$. 
First, they are 
structurally robust, that is, qualitatively invariant under generic changes 
of the substrate potential. Second, the resonant frequencies $\omega=1/n$
mark transitions between IPB and APB behaviour in both the core and tail
senses.

Several caveats should be attached to the picture developed in this section.
First, theorem \ref{th1} gives no guarantee that breathers really will exist
in the whole domain depicted. Generically, it appears that they do, but 
there 
are Klein-Gordon-like systems with smaller existence domains \cite{hasspe}.
Second, we have predicted the decay properties of a breather's Fourier
modes based on the linearized model. Numerically this approximation seems to
work well most of the time. However, occasionally a few modes have 
anomalously
slow spatial decay due to nonlinear corrections \cite{fla}, so figure 
\ref{fig4} may receive fine detail corrections. Third, it may be that some
of the Fourier modes are are wholly absent for symmetry reasons (e.g.\
$V$ is symmetric). The picture is then modified substantially, in much the 
same way as the core picture changes from figure \ref{fig1} to figure
\ref{fig3}. Finally, these results, while plausible, are heuristic. It
would be interesting to seek a rigorous analytic underpinning for them.

\section{Concluding remarks}
\label{con}

In this paper, a systematic correlation between breather initial profiles and their frequencies
was examined. Two types of breather were identified, called IPBs and
APBs,
based on sign correlations of the 3 core lattice sites.
 The numerical data suggest that generically these two types occur in
alternating bands in the $T$ parameter space ($T\in(2\pi,\infty)$), and that
the resonant periods $T\in 2\pi\Z^+$ separate an IPB band from an APB band.
The genericity of this behaviour was proved rigorously.

The distinction between IPBs and APBs, which
 does not appear to have attracted much attention
in the literature, may have phenomenological
implications in applications of the model (\ref{1}). One reason for interest
in discrete breathers (particularly continued one-site breathers)
 is that, because of their strong spatial localization, 
they typically require
little energy to achieve large amplitude oscillations close to the centre.
This makes them good candidates for ``seeds'' of mechanical breakdown of the
network, the idea being that the central oscillation becomes so violent as
to break the chain (a similar mechanism is postulated as a mechanism for
DNA denaturation, for example \cite{dna}). One expects the central intersite
springs to carry a larger proportion of the total breather energy for
APBs than IPBs, making them more energy efficient seeds
of mechanical breakdown. However, further detailed simulations of full
lattice systems are required to test this hypothesis.

\section*{Acknowledgments}

This work was partially completed during a visit by MH to the 
Max-Planck-Institut f\"{u}r Mathematik in den Naturwissenschaften, Leipzig,
where JMS was a guest scientist. Both authors wish to thank Prof.\
Eberhard Zeidler
for the generous hospitality of the institute. JMS was financially supported by an EPSRC Postdoctoral
Research Fellowship in Mathematics. 
 
\section*{Appendix}

We recall some basic definitions and an elementary result of
infinite dimensional differential topology, the Regular Value Theorem, which 
gives conditions under which the level set of a smooth function is 
guaranteed to be a manifold.

\begin{defn}
We say that a closed subspace $S$ of a complete topological vector space 
$B$ {\rm splits} if there
exists another closed subspace $C$ which is complementary to $S$, i.e. 
$C + S = B$ and 
$C \cap S = (0)$.
\end{defn}
Any 
finite dimensional subspace 
(necessarily closed) of a Banach space splits, as does any closed subspace
of finite codimension.

\begin{defn}
  A $C^1$ map $f:X\ra Y$ between Banach manifolds is a {\rm submersion} at 
$x \in X$ if 
$Df_x:T_xX \ra T_{f(x)}Y$ is surjective and ker $Df_x$ splits. A value 
$y \in Y$ is a
{\rm regular value} of $f$ if $f$ is a submersion for every 
$x \in f^{-1}(y)$.
\end{defn}
Note that
in the case that $Y$ is finite dimensional then ker $Df_x$ is of finite 
codimension
in $T_x X$ and hence is guaranteed to split. In this case we need only 
verify that
$Df_x$ is surjective for $f$ to be a submersion at $x$.
\begin{thm}
  (The Regular Value Theorem) Let $f:X \ra Y$ be a $C^1$ map between Banach 
manifolds
and $y \in Y$ be a regular value of $f$. Then $f^{-1}(y)$ is a submanifold 
of $X$ 
with $T_x f^{-1}(y) \cong \mathrm{ker}\  Df_x$.
\end{thm}
The proof follows from working in charts around $x$ and $f(x)$ and applying 
the 
Implicit Function Theorem \cite{chobru2}.

\end{document}